\documentclass[dvipdfm,epsf]{iopart}
\usepackage{iopams,amsfonts,rotating,graphicx,fancybox,fancyhdr,bm,eucal,dsfont,appendix}

\eqnobysec

\begin{document}

\title[Statistics of thermal to shot noise crossover in chaotic cavities]
        {Statistics of thermal to shot noise crossover in chaotic cavities}

\author{Vladimir Al.~Osipov$^1$ and Eugene Kanzieper$^2$}

\address{
    $^1$ Fachbereich Physik, Universit\"at Duisburg-Essen, Duisburg D-47057,
       Germany \\
    $^2$ Department of Applied Mathematics, H.I.T. -- Holon Institute of
    Technology, Holon 58102,
    Israel} \eads{\mailto{vladimir.osipov@uni-due.de} and
    \mailto{eugene.kanzieper@hit.ac.il}}

\begin{abstract}
Recently formulated integrable theory of quantum transport [Osipov
and Kanzieper, Phys. Rev. Lett. {\bf 101}, 176804 (2008)] is
extended to describe sample-to-sample fluctuations of the noise
power in chaotic cavities with broken time-reversal symmetry.
Concentrating on the universal transport regime, we determine
dependence of the noise power cumulants on the temperature, applied
bias voltage, and the number of propagating modes in the leads.
Intrinsic connection between statistics of thermal to shot noise
crossover and statistics of Landauer conductance is revealed and
briefly discussed.
\\ \\
\texttt{arXiv:~0902.3069v2}

\end{abstract}

\tableofcontents

\newpage
\section{Introduction: Thermal versus shot noise}

The charge transfer through a phase-coherent
cavity exhibiting chaotic classical dynamics is a random process
influenced by discreteness of the electron charge $e$ and the
quantum nature of electrons (Blanter and B\"uttiker 2000, Imry 2002, Martin 2005). Fluctuations
of charge transmitted during a fixed time interval or, equivalently,
fluctuations $\delta I(t)$ of current around its mean are quantified
by the noise power
\begin{eqnarray}
    {\mathcal P} = 2\int_{-\infty}^{+\infty} dt\, \left< \delta I(t+t_0) \delta I (t_0)
    \right>_{t_0},
\end{eqnarray}
where the brackets $\left<\cdots \right>_{t_0}$ indicate the time
averaging.

At temperatures $\theta = k_B T$ which are much larger than a bias
voltage ${\upsilon}=eV$ applied to the cavity ($\theta \gg
\upsilon$), the current fluctuations are dominated by the
equilibrium {\it thermal noise}, also known as Johnson-Nyquist
noise. Caused by fluctuating occupation numbers in a flow of
carriers injected into cavity from electronic reservoirs, thermal
noise extends over all frequencies up to the quantum limit
$\theta/h$. In the absence of electron-electron interactions, its
power at zero bias voltage ($\upsilon =0$) is related to the
scattering matrix ${\mathcal S}$ of the system composed of the
cavity and the leads (Khlus 1987, Lesovik 1989, B\"uttiker 1990,
Martin and Landauer 1992, B\"uttiker 1992):
\begin{eqnarray}
\label{pth}
    {\mathcal P}_{\rm th}(\theta) = 4\theta\,G_0\,
    {\rm tr} ({\mathcal C}_1 {\mathcal S} {\mathcal C}_2 {\mathcal S}^\dagger).
\end{eqnarray}
Here, $G_0= e^2/h$ is the conductance quantum. The projection
matrices ${\mathcal C}_{1,2}$ encode the information about
particular cavity-lead geometry and will be specified later on.

In the opposite limit of low temperatures ($\theta \ll \upsilon$),
the current fluctuations are still significant even though the flow
of incident electrons is essentially noiseless. In this temperature
regime, nonequilibrium current fluctuations (known as a {\it shot
noise}) exist because of (i) the granularity of the electron charge
$e$ and (ii) the stochastic nature of electron scattering inside the
cavity which splits the electron wave into two or more partial waves
leaving the cavity through different exits. It is this ``uncertainty
of not knowing where the electron came from and where it will go
to'' (Oberholzer {\it et al} 2002) that makes the transmitted charge
to fluctuate. At zero temperature, the scattering matrix approach
brings the shot noise power in the form
\begin{equation}
\label{pshot}
    {\mathcal P}_{\rm shot}(\upsilon) = 2 \upsilon\,G_0\left[
    {\rm tr} ({\mathcal C}_1 {\mathcal S} {\mathcal C}_2 {\mathcal S}^\dagger)
    - {\rm tr} ({\mathcal C}_1 {\mathcal S} {\mathcal C}_2 {\mathcal S}^\dagger)^2
    \right].
\end{equation}

At finite temperatures, both sources of noise are operative, the
total noise ${\mathcal P}(\theta,\upsilon)$ being a complicated
function of temperature and bias voltage
\footnote{Equation~(\ref{ptotal}) disregards the low-frequency $1/f$
noise that can efficiently be filtered out in experiments.}:
\begin{eqnarray}\fl
\label{ptotal}
    {\mathcal P}(\theta,\upsilon) = 4 \theta \, G_0\,
    \Big(
    {\rm tr} ({\mathcal C}_1 {\mathcal S}
  {\mathcal C}_2 {\mathcal S}^\dagger)^2
    + \frac{\upsilon}{2\theta} \, {\rm coth}
    \left(
        \frac{\upsilon}{2\theta}
    \right)
    \left[
    {\rm tr} ({\mathcal C}_1 {\mathcal S} {\mathcal C}_2 {\mathcal S}^\dagger)
    - {\rm tr} ({\mathcal C}_1 {\mathcal S} {\mathcal C}_2 {\mathcal S}^\dagger)^2
    \right] \Big).
\end{eqnarray}
Equation (\ref{ptotal}) suggests that the crossover from thermal
noise ${\mathcal P}_{\rm th}(\theta)= {\mathcal P}(\theta,0)$ to
shot noise ${\mathcal P}_{\rm shot}(\upsilon)= {\mathcal
P}(0,\upsilon)$ depends in a sensitive way on scattering properties
of the cavity and the leads incorporated in the scattering matrix
${\mathcal S}$. Since chaotic scattering of electrons inside the
cavity induces fluctuations of ${\mathcal S}$-matrix (Bl\"umel and Smilansky 1990),
the noise power ${\mathcal P}(\theta,\upsilon)$ fluctuates, too.

So far, the thermal to shot noise crossover has only been studied at
the level of {\it average} noise power. For the two-terminal
scattering geometry comprised of the cavity attached to outside
reservoirs (kept at temperature $\theta$) via two leads supporting
$N_{\rm L}$ and $N_{\rm R}$ propagating modes, respectively, the
average noise power equals (Blanter and Sukhorukov 2000, Oberholzer
{\it et al} 2001, Savin and Sommers 2006)
\begin{equation}
\label{paverage}
 \left<{\mathcal P}(\theta,\upsilon)\right>_{{\mathcal S}} =
 \left<{\mathcal P}_{\rm th}\right>_{{\mathcal S}}
 \left[
        1 + \frac{N_{\rm L}N_{\rm R}}{(N_{\rm L}+N_{\rm R})^2-1}\, f_\beta
 \right],
\end{equation}
where
\begin{equation}
\label{peq}
 \left<{\mathcal P}_{\rm th}\right>_{{\mathcal S}} =
 4\theta\, G_0 \frac{N_{\rm L} N_{\rm R}}{N_{\rm L}+N_{\rm R}}
\end{equation}
is the average equilibrium thermal noise power, and the
thermodynamic function
\begin{eqnarray}
\label{fbeta}
    f_\beta = \beta \coth \beta -1
\end{eqnarray}
depends on the ratio $\beta=\upsilon/2\theta$ between the bias
voltage $\upsilon$ and the temperature $\theta$. Equations
(\ref{paverage}) and (\ref{peq}) hold for cavities with broken time
reversal symmetry \footnote{The two can readily be extended to other
symmetry classes, see Savin and Sommers (2006).}. Derived for the
universal transport regime (Beenakker 1997, Richter and Sieber 2002,
M\"uller {\it et al} 2007) emerging in the limit $\tau_{\rm D} \gg
\tau_{\rm E}$ (Agam {\it et al} 2000), where $\tau_{\rm D}$ is the
average electron dwell time and $\tau_{\rm E}$ is the Ehrenfest time
(the time scale where quantum effects set in), the above prediction
has been confirmed in a remarkable series of experiments (Oberholzer
{\it et al} 2001, Cron{\it et al} 2001, Oberholzer {\it et al}
2002).

In this paper, we examine {\it statistics} of the thermal to shot
noise crossover. The latter, contained in the distribution function
of the noise power ${\mathcal P}(\theta,\upsilon)$ or, equivalently,
in its {\it cumulants} $\langle\!\langle {\mathcal P}^\ell
\rangle\!\rangle$, can effectively be described within the framework
of integrable theory of quantum transport formulated by Osipov and
Kanzieper (2008). Let us stress that recent experimental studies
(Flindt {\it et al} 2009) of quantum noise fluctuations in nanoscale
conductors (which concentrated on detection of higher cumulants of
noise) suggest that testing our predictions may be feasible within
the current limits of nanotechnology.

\section{Integrable theory of noise power fluctuations}

In what follows, we consider chaotic cavities with broken
time-reversal symmetry which are probed, via ballistic point
contacts, by two (left and right) leads; the leads supporting
$N_{\rm L}$ and $N_{\rm R}$ propagating modes, respectively, are
further coupled to external reservoirs kept at the temperature
$\theta$. This scattering geometry corresponds to the projection
matrices ${\mathcal C}_{1,2}$ of the form
\begin{eqnarray}
{\mathcal C}_1= \left(
                  \begin{array}{cc}
                    \mathds{1}_{N_{\rm L}} & 0 \\
                    0 & 0_{N_{\rm R}} \\
                  \end{array}
                \right),\;\;\;
{\mathcal C}_2= \left(
                  \begin{array}{cc}
                    0_{N_{\rm L}} & 0 \\
                    0 & \mathds{1}_{N_{\rm R}} \\
                  \end{array}
                \right),
\end{eqnarray}
see Eqs.~(\ref{pth}) -- (\ref{ptotal}).

\subsection{Joint cumulants of Landauer conductance and noise power}
The starting point of our analysis is the {\it joint} cumulant
generating function (JCGF)
\begin{eqnarray}
\label{cgf-def}
    {\mathcal F}_{n}(z,w) = \left<
        \exp(-z \,G/G_0)\exp(- w\, {\mathcal P}/{\mathcal P}_0)
    \right>_{{\mathcal S}\in {\rm CUE}(N)}
\end{eqnarray}
of the Landauer conductance $G = G_0 {\rm tr} ({\mathcal C}_1
{\mathcal S} {\mathcal C}_2 {\mathcal S}^\dagger)$ and the noise
power ${\mathcal P}(\theta,\upsilon)$ measured in the units of $G_0=
e^2/h$ and ${\mathcal P}_0 =  4\theta\, G_0$, respectively. The
joint dimensionless cumulants
\begin{eqnarray}
\kappa_{\ell,m} =
    \langle\!\langle (G/G_0)^\ell ({\mathcal P}/{\mathcal P}_0)^m
    \rangle\!\rangle
\end{eqnarray}
can be extracted from the expansion
\begin{equation}
\label{c-def}
    \log {\mathcal F}_n(z,w) = \sum_{\ell,m=0}^\infty (-1)^{\ell+m} \frac{z^\ell w^m}{\ell!\,m!}\,
    \kappa_{\ell,m},
\end{equation}
where $\kappa_{0,0}\equiv 0$. In both Eqs.~(\ref{cgf-def}) and
(\ref{c-def}), the subscript $n$ stands for\linebreak $n=\min(N_{\rm
L},N_{\rm R})$, and $N=N_{\rm L}+N_{\rm R}$ is the total number of
propagating modes (channels) in the leads. The notation ${\mathcal
S} \in {\rm CUE}(N)$ indicates that averaging runs over scattering
matrices ${\mathcal S}$ drawn from the Dyson circular unitary
ensemble (Bl\"umel and Smilansky 1990, Mello and Baranger 1999,
Mehta 2004). The latter is microscopically justified (Lewenkopf and
Weidenm\"uller 1991, Brouwer 1995) in the universal transport regime
we are confined to.

To perform the averaging in Eq.~(\ref{cgf-def}) in a most economic
way, we employ a polar decomposition (Hua 1963, Baranger and Mello
1994, Forrester 2006) of ${\mathcal S}$-matrix. Bringing into play a
set of $n$ transmission eigenvalues ${\bm T} = (T_1,\cdots,T_n) \in
(0,1)^n$ distributed in accordance with the joint probability
density function
\begin{eqnarray}
\label{PnT}
    P_n({\bm T}) = c_n^{-1} \Delta_n^2({\bm T}) \prod_{j=1}^n
    T_j^\nu,
\end{eqnarray}
this decomposition highlights Landauer's idea of viewing conductance
as transmission,
\begin{eqnarray}
G({\bm T}) = G_0\sum_{j=1}^n T_j.
\end{eqnarray}
Simultaneously, it reduces the expression for noise power
[Eq.~(\ref{ptotal})] down to
\begin{eqnarray}
\label{ptotal-T}
    {\mathcal P}({\bm T})
    ={\mathcal P}_0 \left(
    \sum_{j=1}^n T_j + f_\beta \sum_{j=1}^n T_j(1-T_j)\right).
\end{eqnarray}
The parameter $\nu$ in Eq.~(\ref{PnT}) is a measure of asymmetry
between the leads,\linebreak $\nu = |N_{\rm L}-N_{\rm R}|$, the
notation $\Delta_n({\bm T})$ stands for the Vandermonde determinant
$\Delta_n ({\bm T})  =\prod_{j<k} (T_k-T_j)$, whilst $c_n$ is a
normalisation constant. As the result, we are left with the JCGF in
the form
\begin{equation}
\label{Fnzw}
    {\mathcal F}_n(z,w) = c_n^{-1}\int_{(0,1)^n} \prod_{j=1}^n dT_j\,T_j^\nu \,\Gamma_{z,w}(T_j)
    \,\Delta_n^2({\bm T}),
\end{equation}
where
\begin{eqnarray}
\label{Gamma}
    \Gamma_{z,w}(T) =
    \exp\left[
        -(z+w)\,T - w\,f_\beta\, T(1-T)
    \right].
\end{eqnarray}
Although the above matrix integral representation of the JCGF
${\mathcal F}_n(z,w)$ is by far more complicated than the one
appearing in the integrable theory of conductance fluctuations
(Osipov and Kanzieper 2008),
\begin{eqnarray} \fl \label{ccgf}
    {\mathcal F}_n(z,0) =
    \left<
        \exp(-z \,G/G_0)\right>_{{\mathcal S}\in {\rm CUE}(N)}
        =
        c_n^{-1}\int_{(0,1)^n} \prod_{j=1}^n dT_j\,T_j^\nu \,e^{-z T_j}
    \,\Delta_n^2({\bm T}),
\end{eqnarray}
it can still be treated nonperturbatively, much in line with the
formalism used in the exact approach to zero-dimensional replica
sigma models (Kanzieper 2002, Kanzieper 2005, Osipov and Kanzieper
2007, Kanzieper 2009).

\subsection{The $\tau$ function theory of the joint cumulant generating function}

The ``deform-and-study'' approach (Morozov 1994, Adler {\it et al}
1995, Adler and van Moerbeke 2001) borrowed from the theory of
integrable systems is central to the nonperturbative calculation of
${\mathcal F}_n(z,w)$. In the present context, the main idea of the
method consists of ``embedding'' ${\mathcal F}_n(z,w)$ into a more
general theory of the $\tau$ function
\begin{equation}
    \label{tau-def}
    \tau_n({\bm t};z,w) = \frac{1}{n!}\int_{(0,1)^n} \prod_{j=1}^n dT_j\,T_j^\nu \,\Gamma_{z,w}(T_j)
    \,e^{V({\bm t};T_j)} \Delta_n^2({\bm T})
\end{equation}
which possesses the infinite-dimensional parameter space ${\bm
t}=(t_1,t_2,\dots)$ arising as the result of the ${\bm t}$
deformation
\begin{eqnarray}
V({\bm t};T) = \sum_{k=1}^\infty t_k T^k.
\end{eqnarray}
Studying an evolution of the $\tau$ function in the extended $(n,
{\bm t},z,w)$ space allows us to identify various nonlinear
differential hierarchical relations. A projection of these relations
onto the hyperplane ${\bm t}={\bm 0}$,
\begin{eqnarray}
\label{proj}
    {\mathcal F}_n(z,w) = \frac{n!}{c_n}\,
    \tau_n({\bm t};z,w)\Big|_{{\bm t}={\bm 0}},
\end{eqnarray}
generates, among others, a closed nonlinear differential equation
for the JCGF ${\mathcal F}_n(z,w)$. It is this equation
[Eq.~(\ref{ndeq})] that will further supply the cumulants of noise
power.

The two key ingredients of the exact theory of $\tau$ functions are
(i) the bilinear identity (Date {\it et al} 1983) and (ii) the
(linear) Virasoro constraints (Mironov and Morozov 1990).

\subsubsection{Bilinear identity and Kadomtsev-Petviashvili equation}\noindent\newline\newline
The bilinear identity encodes an infinite set of hierarchically
structured nonlinear differential equations in the variables ${\bm
t}=(t_1, t_2,\dots)$. For the model introduced in
Eq.~(\ref{tau-def}), the bilinear identity reads (Adler {\it et al}
1995, Tu {\it et al} 1996):
\begin{eqnarray} \label{bi-id}\fl
    \oint_{{\cal C}_\infty} dz \,e^{a\, v(\bm{t-t^\prime};z)}
    \tau_{n}(\bm{t}-[\bm{z}^{-1}])\,
    \frac{\tau_{m+1}(\bm{t^\prime}+[\bm{z}^{-1}])}{z^{m+1-n}}
    \nonumber\\
    =\oint_{{\cal C}_\infty} dz \,e^{(a-1)\, v(\bm{t-t^\prime};z)}
    \tau_m(\bm{t^\prime}-[\bm{z}^{-1}]) \frac{\tau_{n+1}
    (\bm{t}+[\bm{z}^{-1}])}{z^{n+1-m}}.
\end{eqnarray}
Here, $a\in {\mathbb R}$ is a free parameter; the integration
contour ${\cal C}_\infty$ encompasses the point $z=\infty$; the
notation ${\bm t} \pm [{\bm z}^{-1}]$ stands for the infinite set of
parameters $\{t_j\pm z^{-j}/j\}$; for brevity, both $z$ and $w$ were
dropped from the arguments of $\tau$ functions.

Being expanded in terms of $\bm{ t^\prime}-{\bm t}$ and $a$,
Eq.~(\ref{bi-id}) generates a variety of integrable hierarchies
(Osipov and Kanzieper 2009). One of them is the
Kadomtsev-Petviashvili (KP) hierarchy. Its first nontrivial member,
\begin{eqnarray} \fl
    \label{fkp}
        \left(
    \frac{\partial^4}{\partial t_1^4} + 3\,\frac{\partial^2}{\partial t_2^2} -
        4\, \frac{\partial^2}{\partial t_1 \partial t_3}
    \right)\, \log \tau_n({\bm t};z,w)
    + \,6\, \left(
        \frac{\partial^2}{\partial t_1^2}\, \log \tau_n({\bm t};z,w)
    \right)^2 = 0
\end{eqnarray}
is of primary importance since its projection onto ${\bm t}={\bm 0}$
[Eq.~(\ref{proj})] gives rise to a nonlinear differential equation
for the JCGF ${\mathcal F}_n(z,w)$. The resulting equation will
further be used to determine the noise power cumulants we are aimed
at.

\subsubsection{Virasoro constraints}\noindent\newline\newline
Since we are interested in deriving a differential equation for
${\mathcal F}_n(z,w)$ in terms of the derivatives over variables $z$
and $w$, we have to seek an additional block of the theory that
would make a link between $t_j$-derivatives in Eq.~(\ref{fkp}) taken
at ${\bm t}={\bm 0}$ and the derivatives over $w$ and $z$. This
missing block is the {\it Virasoro constraints} which reflect the
invariance of the $\tau$ function [Eq.~(\ref{tau-def})] under a
change of the integration variables.

In the present context, it is useful to demand the invariance under
the set of transformations
\begin{eqnarray}
\label{v-tr}
T_j \rightarrow \tilde{T}_j +
    \epsilon \,\tilde{T}_j^{q+1} (\tilde{T}_j-1),\;\;\; q\ge 0.
\end{eqnarray}
Employing by now a standard procedure (Mironov and Morozov 1990,
Adler and van Moerbeke 1995), one readily checks that the
transformation (\ref{v-tr}) induces Virasoro constraints in the form
\begin{equation}
\label{vc-1}
    \big[ \hat{L}_{q+1}({\bm t}) - \hat{L}_{q}({\bm t})\big] \tau_n({\bm t};z,w)
     = 0,\;\;\;  q\ge 0,
\end{equation}
where a set of differential operators
\begin{eqnarray}\fl
\label{vc-2}
    \hat{L}_q({\bm t})
     =
     \hat{\mathcal L}_{q}({\bm t}) + 2f_\beta\, w
        \frac{\partial}{\partial t_{q+2}} - [z+(1+f_\beta)\,w] \frac{\partial}{\partial t_{q+1}}
        + \nu \frac{\partial}{\partial t_q}
\end{eqnarray}
involves the Virasoro operators
\begin{eqnarray}
    \label{vc-3}
    \hat{{\cal L}}_q({\bm t}) = \sum_{j=1}^\infty jt_j \,\frac{\partial}{\partial t_{q+j}}
    +
    \sum_{j=0}^q \frac{\partial^2}{\partial t_j \partial t_{q-j}},
\end{eqnarray}
satisfying the Virasoro algebra
\begin{eqnarray}
    [\hat{{\cal L}}_p,\hat{{\cal L}}_q] = (p-q)\hat{{\cal
    L}}_{p+q}, \;\;\;\; p,q\ge -1.
\end{eqnarray}
In Eqs.~(\ref{vc-2}) and (\ref{vc-3}), the convention
$\partial/\partial t_0 \equiv n$ is assumed.

\subsubsection{Nonlinear differential equation for ${\mathcal F}_n(z,w)$}\label{s-223}\noindent\newline\newline
To derive a differential equation for the JCGF ${\mathcal
F}_n(z,w)$, one has to project the first KP equation Eq.~(\ref{fkp})
onto the hyperplane ${\bm t}={\bm 0}$. Spotting the identities
($f_\beta>0$)
\begin{eqnarray}
\frac{\partial}{\partial t_1} \tau_n({\bm t};z,w)&=& - \frac{\partial}{\partial z}  \tau_n({\bm t};z,w),\\
f_\beta \frac{\partial}{\partial t_2}  \tau_n({\bm t};z,w) &=& \frac{\partial}{\partial w}  \tau_n({\bm t};z,w)
- (1+f_\beta)\frac{\partial}
{\partial z}  \tau_n({\bm t};z,w),
\end{eqnarray}
we combine Eqs.~(\ref{fkp}) with the Virasoro constraints
Eq.~(\ref{vc-1}) taken at $q=0$,
\begin{eqnarray} \fl
    \Bigg[ \sum_{j=1}^\infty jt_j \left(
    \frac{\partial}{\partial t_{j+1}}
    -
    \frac{\partial}{\partial t_{j}}
    \right)
    +  2f_\beta\, w \left(
        \frac{\partial}{\partial t_{3}} -
        \frac{\partial}{\partial t_{2}}
    \right) \nonumber \\ \fl
         \quad - \left[z+(1+f_\beta)\,w\right] \left(
     \frac{\partial}{\partial t_{2}}
     -
     \frac{\partial}{\partial t_{1}}
     \right)
     + (N_{\rm L}+N_{\rm R}) \frac{\partial}{\partial t_1}
     -N_{\rm L}N_{\rm R}\Bigg]\,  \tau_n({\bm t};z,w)=0,
\end{eqnarray}
to derive:
\begin{eqnarray}\fl
\label{ndeq}
    \Bigg[ w f_\beta^2 \frac{\partial^4}{\partial z^4} + \left[2(N_{\rm L}+N_{\rm R})f_\beta -  2 z + w\,(1-f_\beta^2)\right]
    \frac{\partial^2}{\partial z^2} + 2(z-2w)\frac{\partial^2}{\partial z \partial w}
    \nonumber \\  \fl
     \quad + 3 w \frac{\partial^2}{\partial w^2} + 2 \left(\frac{\partial}{\partial w} -  \frac{\partial}{\partial z}\right) \Bigg]
    \,\log {\mathcal F}_n(z,w)+ 6 w\,f_\beta^2  \left(
        \frac{\partial^2}{\partial z^2} \, \log {\mathcal F}_n(z,w)
    \right)^2
      = 0.
\end{eqnarray}
Owing to Eq.~(\ref{c-def}), this nonlinear equation considered {\it
together} with the equation for ${\mathcal F}_n(z,0)$ contains all
the information about joint cumulants of the Landauer conductance
and the noise power. The latter equation, written in terms of
\begin{eqnarray}
\label{sF}
    \sigma_n(z) = N_{\rm L} N_{\rm R} + z \frac{\partial}{\partial z}\log {\mathcal F}_n(z,0),
\end{eqnarray}
reads (Osipov and Kanzieper 2008):
\begin{eqnarray}
\label{OK-2008}\fl
    z^2 \frac{\partial^3}{\partial z^3} \sigma_n(z) +
    z \frac{\partial^2}{\partial z^2} \sigma_n(z) +
    6 z \left(
        \frac{\partial}{\partial z} \sigma_n(z)
    \right)^2
    - 4 \sigma_n \frac{\partial}{\partial z} \sigma_n(z) \nonumber \\
    \fl \qquad-
    \left[
        \Big(
            z -(N_{\rm L}+ N_{\rm R})
        \Big)^2 - 4 N_{\rm L}N_{\rm R}
    \right] \frac{\partial}{\partial z} \sigma_n(z) -
    (N_{\rm L}+ N_{\rm R} -z) \sigma_n(z) = 0.
\end{eqnarray}
This can be recognised as the Chazy form (Chazy 1911, Cosgrove and
Scoufis 1993) of the fifth Painlev\'e transcendent
\begin{eqnarray}
\label{s5-chazy} \fl
    \left(z \frac{\partial^2}{\partial z^2}
 \sigma_n(z)\right)^2 - \left[\sigma_n(z) +
 2 \left(\frac{\partial}{\partial z}\sigma_n(z)\right)^2 + (N_{\rm L}+N_{\rm R}-z)\frac{\partial}{\partial z}\sigma_n(z)
 \right]^2 \nonumber\\
\fl \qquad\qquad
+ 4\left(\frac{\partial}{\partial z}\sigma_n(z)\right)^2
    \left(N_{\rm L}+\frac{\partial}{\partial z}\sigma_n(z)\right)
    \left(N_{\rm R}+\frac{\partial}{\partial z}\sigma_n(z)\right)
    =0
\end{eqnarray}
written in the Jimbo-MIwa-Okamoto form (Jimbo {\it et al} 1980,
Okamoto 1987). For completeness, we have included a detailed
derivation of Eq.~(\ref{OK-2008}) into Appendix A.

\subsection{Recurrence solution for joint cumulants}
Combined with the cumulant expansion Eq.~(\ref{c-def}), the
differential equation Eq.~(\ref{ndeq}) furnishes the nonlinear
recurrence for the joint dimensionless cumulants $\kappa_{\ell,m}$
of conductance and noise power ($\ell,m \ge 0$):
\begin{eqnarray}
\label{2drec} \fl
    m\,\Big[f_\beta^2\, \kappa_{\ell+4,m-1} + (1-f_\beta^2) \,\kappa_{\ell+2,m-1}\Big] - 2 \,
    (N_{\rm L}+N_{\rm R})\,  f_\beta \,\kappa_{\ell+2,m} \nonumber \\
    \fl \qquad\qquad\quad
    - 2\left(\ell+ 2 m + 1 \right) \, \kappa_{\ell+1,m}    +  (2\ell+3m+2) \, \kappa_{\ell,m+1}
    \nonumber\\
    \fl \qquad\qquad\quad
     + \, 6m \,f_\beta^2
    \sum_{i=0}^{m-1}\left( {m-1}\atop{i} \right) \sum_{j=0}^\ell   \left( {\ell}\atop{j} \right)
    \kappa_{j+2,i}\, \kappa_{\ell-j+2,m-i-1}=0.
\end{eqnarray}
To resolve it, one must know the boundary conditions whose r\^ole is
played by cumulants $\kappa_{\ell,0} =\langle \! \langle
(G/G_0)^\ell \rangle\!\rangle$ of the dimensionless Landauer
conductance. These have been nonperturbatively calculated in our
previous publication (Osipov and Kanzieper 2008). Indeed, given the
mean conductance
\begin{eqnarray}
\label{k10}
    \kappa_{1,0} = \frac{N_{\rm L} N_{\rm R}}{N_{\rm L} + N_{\rm R}}
\end{eqnarray}
and its variance
\begin{eqnarray}
\label{k20}
    \kappa_{2,0} = \frac{\kappa_{1,0}^2}{(N_{\rm L} + N_{\rm R})^2-1},
\end{eqnarray}
the higher order cumulants $\kappa_{\ell,0}$'s are determined by the
one-dimensional recurrence~\footnote{Equations (\ref{k10}),
(\ref{k20}) and (\ref{cumeq}) follow from the cumulant expansion
\begin{eqnarray}
    \log {\mathcal F}_n(z,0) = \sum_{\ell=1}^\infty (-1)^\ell \frac{z^\ell}{\ell!}
    \kappa_{\ell,0}, \nonumber
\end{eqnarray}
see Eqs.~(\ref{c-def}) and (\ref{ccgf}), substituted into
Eq.~(\ref{s5-chazy}). Successive iterations of Eq.~(\ref{cumeq})
yield the cumulants $\kappa_{\ell,0}$ of Landauer conductance in the
form
\begin{eqnarray}
    \kappa_{\ell,0} = \frac{(\ell-1)!}{\prod_{j=1}^{\ell-1} (N^2-j^2)}\, p_\ell(\kappa_{1,0}), \nonumber
\end{eqnarray}
where the first few polynomials $p_\ell(\kappa)$ are:
\begin{eqnarray}
    p_1(\kappa) =\kappa, \nonumber\\
    p_2(\kappa) = \kappa^2, \nonumber\\
    p_3(\kappa) = 4 \kappa^3 - N\kappa^2, \nonumber\\
    p_4(\kappa) = 12\left(2 - \frac{1}{N^2-1}\right) \kappa^4 - 10 N\kappa^3 + (N^2+1)\kappa^2.\nonumber
\end{eqnarray}
Here, $N=N_{\rm L}+N_{\rm R}$.}
\begin{eqnarray}
\label{cumeq}
\fl [(N_{\rm L}+N_{\rm R})^2-\ell^2]\,(\ell+1) \kappa_{\ell+1,0}
        + (N_{\rm L}+N_{\rm R}) (2\ell-1)\,\ell \kappa_{\ell,0} \nonumber\\
        \fl \qquad + \ell(\ell-1)(\ell-2)\,
     \kappa_{\ell-1,0}
     - 2 \sum_{j=0}^{\ell-1} (3j+1) (j-\ell)^2 \left({\ell}\atop{j}\right)
     \kappa_{j+1,0}\, \kappa_{\ell-j,0} = 0.
\end{eqnarray}

Equations (\ref{2drec}) and (\ref{cumeq}) represent the main result
of our study~\footnote{Notice that Eq.~(15) in the paper by Osipov
and Kanzieper (2008) contains typos. The correct formula is given by
Eq.~(\ref{cumeq})}. They provide a nonperturbative description of
the noise power fluctuations in the crossover region between the
thermal and the shot noise (Savin, Sommers and Wieczorek 2008)
regimes (as discussed in the Introduction) {\it by relating the
temperature ($\theta$) and bias-voltage ($\upsilon$) dependent
cumulants of the noise power to those of the Landauer conductance}.

Undoubtedly, the very existence of the above nontrivial relation
(which emphasises a fundamental r\^{o}le played by Landauer
conductance in transport problems) must be well rooted in the
mathematical formalism and also have a good physics reason. As far
as the former point is concerned, we wish to stress that a na\"{i}ve
attempt to build a theory for the generating function ${\mathcal
F}_n(0,w)$ of solely noise power cumulants faces an unsurmountable
obstacle: the KP equation [Eq.~(\ref{fkp})] and appropriate Virasoro
constraints [Eq.~(\ref{vc-1}) at $z=0$] cannot be resolved jointly
in the hyperplane ${\bm t}={\bm 0}$. This justifies the starting
point [Eq.~(\ref{cgf-def})] of our analysis. The physics arguments
behind the peculiar structure of our solution are yet to be found.

\subsection{Noise power cumulants in the crossover regime}
Some computational effort is needed to read off explicit formulae
for the noise power cumulants from Eqs.~(\ref{2drec}) and
(\ref{cumeq}). Below we provide expressions for two families of
joint cumulants expressed in terms of dimensionless cumulants
$\kappa_{\ell,0} =\langle \! \langle (G/G_0)^\ell \rangle\!\rangle$
of the Landauer conductance.
\begin{itemize}
  \item Mean noise power
        \begin{eqnarray} \label{mnp}
        \langle\!\langle{\mathcal P} \rangle\!\rangle = 4\theta\, G_0
        \Big[ N f_\beta\,
        \kappa_{2,0} + \kappa_{1,0} \Big]
        \end{eqnarray}
        is generated by the lowest order member $(\ell,m)=(0,0)$
        of the recurrence Eq.~(\ref{2drec}). Being in concert
        with the known expression Eqs. (\ref{paverage}) and
        (\ref{peq}), this result is a particular case of a more
        general formula
        \begin{eqnarray}
        \label{k-l-1}
            \kappa_{\ell,1} = \kappa_{\ell+1,0} + N\frac{f_\beta}{\ell+1}\, \kappa_{\ell+2,0}.
        \end{eqnarray}
  \item Noise power variance,
        \begin{eqnarray}
 \label{p-cum-2}\fl
 \langle\!\langle{\mathcal P}^2 \rangle\!\rangle = (4\theta\, G_0)^2
 \Bigg[
    \left(
        \frac{2}{3}N^2-1
    \right) \frac{f_\beta^2}{5} \,\kappa_{4,0} + N f_\beta\, \kappa_{3,0}+\left(
        1+ \frac{f_\beta^2}{5}
    \right)\, \kappa_{2,0} - \frac{6}{5} f_\beta^2 \kappa_{2,0}^2
 \Bigg], \nonumber\\
 {}
\end{eqnarray}
is supplied by the $(\ell,m)=(0,1)$ member of the recurrence.
Its generalisation reads:
\begin{eqnarray} \fl \label{k-l-2}
    \kappa_{\ell,2} = \left(
        \frac{2N^2}{\ell+3}-1
    \right) \frac{f_\beta^2}{2\ell+5}\kappa_{\ell+4,0} + 2N \frac{f_\beta}{\ell+2}\kappa_{\ell+3,0}
    +\left(
        1+ \frac{f_\beta^2}{2\ell+5}
    \right) \kappa_{\ell+2,0} \nonumber\\
    - 6 \frac{f_\beta^2}{2\ell+5} \sum_{j=0}^\ell \left({\ell}\atop{j}\right) \kappa_{j+2,0}\kappa_{\ell+2-j,0}.
\end{eqnarray}
Here and above, $N=N_{\rm L}+N_{\rm R}$.
\end{itemize}
\noindent\newline The noise power cumulants $\langle\!\langle
{\mathcal P}^\ell \rangle\!\rangle$ of higher order ($\ell \ge 3$)
can be calculated in the same manner albeit explicit expressions
become increasingly cumbersome. Varying therein the parameters
$(\theta,\upsilon)$ from $(\theta,0)$ to $(0,\upsilon)$, one
observes a smooth crossover between the thermal and the shot noise
regime.

\subsection{Large-$n$ analysis of joint cumulants: Symmetric leads}
The nonpeturbative solution Eq.~(\ref{2drec}) and (\ref{cumeq}) has
a drawback: it does not provide much desired {\it explicit}
dependence of conductance and/or noise power cumulants
$\kappa_{\ell,m}$ on parameters of the scattering system. To probe
such a dependence, we turn to the large-$n$ limit of the recurrence
Eq.~(\ref{2drec}). In what follows, the asymmetry parameter $\nu$
will be set to zero.

Under the latter assumption ($\nu=0$), the joint cumulants
$\kappa_{\ell,m}$ are solutions to the recurrence equation ($\ell,m
\ge 0$)
\begin{eqnarray}
\label{2drec=0} \fl
    m\,\Big[f_\beta^2\, \kappa_{\ell+4,m-1} + (1-f_\beta^2) \,\kappa_{\ell+2,m-1}\Big] - 4n\,
     f_\beta \,\kappa_{\ell+2,m} \nonumber \\
    \fl \qquad\qquad\quad
    - 2\left(\ell+ 2 m + 1 \right) \, \kappa_{\ell+1,m}    +  (2\ell+3m+2) \, \kappa_{\ell,m+1}
    \nonumber\\
    \fl \qquad\qquad\quad
     + \, 6m \,f_\beta^2
    \sum_{i=0}^{m-1}\left( {m-1}\atop{i} \right) \sum_{j=0}^\ell   \left( {\ell}\atop{j} \right)
    \kappa_{j+2,i}\, \kappa_{\ell-j+2,m-i-1}=0
\end{eqnarray}
which must be supplemented by yet another recurrence ($\ell \ge 2$)
\begin{eqnarray}
\label{cumeq=0}
\fl (4n^2-\ell^2)\,(\ell+1) \kappa_{\ell+1,0}
        + 2n (2\ell-1)\,\ell \kappa_{\ell,0} \nonumber\\
        \fl \qquad + \ell(\ell-1)(\ell-2)\,
     \kappa_{\ell-1,0}
     - 2 \sum_{j=0}^{\ell-1} (3j+1) (j-\ell)^2\left({\ell}\atop{j}\right)
     \kappa_{j+1,0}\, \kappa_{\ell-j,0} = 0
\end{eqnarray}
that brings, in turn, a set of initial conditions $\kappa_{\ell,0}$
to Eq.~(\ref{2drec=0}).

\subsubsection{Cumulants of Landauer conductance}\label{s-251}\noindent\newline\newline
It is instructive to start with the asymptotic analysis of
Eq.~(\ref{cumeq=0}). In case of symmetric leads, the conductance
cumulants of odd order vanish~\footnote{At the formal level, this is
direct consequence of the identity ${\mathcal F}_n(z,0) = e^{-nz}
{\mathcal F}_n(-z,0)$ holding as soon as $\nu=0$, see
Eq.~(\ref{ccgf}).}, $\kappa_{2\ell+1,0}\equiv 0$ for all $\ell\ge 1$
albeit $\kappa_{1,0}=n/2$. As the result, one is left with the
recurrence equation for the cumulants $\kappa_{2\ell,0}$ of even
order ($\ell\ge 1$)
\begin{eqnarray}
\label{cumeq=0=even}
\fl [4n^2-(2\ell+1)^2]\,(\ell+1) \kappa_{2\ell+2,0}
       + \ell(4\ell^2-1)\,
     \kappa_{2\ell,0} \nonumber\\
     - 8 \sum_{j=0}^{\ell-1} (3j+2) (j-\ell)^2 \left({2\ell+1}\atop{2j+1}\right)
     \kappa_{2j+2,0}\, \kappa_{2\ell-2j,0} = 0
\end{eqnarray}
subject to the initial condition [Eq.~(\ref{k20})]
\begin{eqnarray}
\label{k20exp}
    \kappa_{2,0} = \frac{n^2}{4(4n^2-1)} = \frac{1}{16}\sum_{\sigma=0}^\infty \frac{1}{(4n^2)^\sigma}.
\end{eqnarray}
Since, in the limit of a large number of propagating modes ($n\gg
1$), the conductance distribution is expected to roughly follow the
Gaussian law (Politzer 1989) with the mean $\kappa_{1,0}^{{\rm
(G)}}=n/2$ and the variance $\kappa_{2,0}^{(\rm G)}=1/16$ [see
Eqs.~(\ref{k10}) and (\ref{k20})], it is natural to seek a large-$n$
solution to Eq.~(\ref{cumeq=0=even}) in the form ($j \ge 1$)
\begin{eqnarray}
\kappa_{2\ell,0} = \frac{1}{16}\delta_{\ell,1} + \delta\kappa_{2\ell,0},
\end{eqnarray}
where $\delta\kappa_{2\ell,0}$ (with $\ell\ge 2$) account for
deviations from the Gaussian distribution. Putting forward the
large-$n$ ansatz
\begin{eqnarray}
\label{1-over-n}
\delta\kappa_{2\ell,0} =
\frac{1}{n^{2\ell}}\sum_{\sigma=0}^\infty\frac{a_{2\ell}(2\sigma)}{n^{2\sigma}}, \;\; \ell\ge 1,
\end{eqnarray}
where [see Eq.~(\ref{k20exp})]
\begin{eqnarray}
\label{a2m}
    a_2 (2\sigma) = \frac{1}{2^{2\sigma+4}},
\end{eqnarray}
we further substitute it into Eq.~(\ref{cumeq=0=even}) to derive:
\begin{eqnarray}
    \frac{a_{2\ell+2}(0)}{a_{2\ell}(0)} = \frac{1}{2^4} \frac{(2\ell+1)!}{(2\ell-1)!},\;\;\; \ell\ge 2,
\end{eqnarray}
and
\begin{eqnarray}
    \frac{a_4(0)}{a_2(2)} = \frac{3!}{2^4}.
\end{eqnarray}
Hence,
\begin{eqnarray}
\label{a2l0}
    a_{2\ell}(0) = \frac{1}{4} \frac{(2\ell-1)!}{4^{2\ell}}.
\end{eqnarray}
Taken together with Eq.~(\ref{a2m}), this yields the leading term in
the $1/n$ expansion [Eq.~(\ref{1-over-n})] for conductance
cumulants:
\begin{eqnarray}\label{cum-res}
    \delta\kappa_{2\ell} \simeq \frac{1}{4}\frac{(2\ell-1)!}{(4n)^{2\ell}}.
\end{eqnarray}
The higher order corrections to Eq.~(\ref{cum-res}) can be obtained
in a regular way.

\subsubsection{Dependence of the noise power cumulants on temperature and bias voltage}
\label{s-252}\noindent\newline\newline Similarly to the previous
subsection, we start an asymptotic analysis of the recurrence
Eq.~(\ref{2drec=0}) with singling out the large-$n$ Gaussian part:
\begin{eqnarray}\fl \label{klm-gauss}
    \kappa_{\ell,m} = \frac{n}{2} \left[
    \delta_{\ell,1}\delta_{m,0} + \left(1+\frac{f_\beta}{4}\right)\delta_{\ell,0}\delta_{m,1}
    \right]\nonumber\\
    + \frac{1}{16}
    \left[\delta_{\ell,1}\delta_{m,1} +
        \delta_{\ell,2}\delta_{m,0}  + \left(
            1+ \frac{f_\beta^2}{8}
        \right)\delta_{\ell,0}\delta_{m,2}
    \right] +\delta\kappa_{\ell,m}.
\end{eqnarray}
The Gaussian part was read off from Eqs.~(\ref{k-l-1}) and
(\ref{k-l-2}); the term $\delta\kappa_{\ell,m}$ accommodates
non-Gaussian corrections to the joint cumulants of Landauer
conductance and the noise power. Their large-$n$ behavior can be
studied within the $1/n$ ansatz
\begin{eqnarray}
\label{ans-joint}
    \delta\kappa_{\ell,m} = \frac{1}{n^{\ell+m}}\sum_{\sigma=0}^\infty \frac{a_{\ell,m}(\sigma)}{n^\sigma},
\end{eqnarray}
where $a_{1,0}(\sigma)=0$. Substitution of Eqs.~(\ref{klm-gauss})
and (\ref{ans-joint}) into the two-dimensional recurrence
Eq.~(\ref{2drec=0}) brings the recurrence equation ($\ell+m>0$)
\begin{eqnarray} \label{rr-a}\fl
    m\left(
        1 - \frac{f_\beta^2}{4}
    \right)\, a_{\ell+2,m-1}(0) - 4 f_\beta a_{\ell+2,m}(0) \nonumber\\
        -2 (\ell+1+2m) a_{\ell+1,m}(0) + (2\ell+2+3m)\,a_{\ell,m+1}(0)=0
\end{eqnarray}
for the expansion coefficients $a_{\ell,m}(0)$ appearing in
Eq.~(\ref{ans-joint}). The (unique) solution of Eq.~(\ref{rr-a}),
subject to the boundary condition
\begin{eqnarray}
    a_{\ell,0}(0) =\frac{(\ell-1)!}{2^{2\ell+3}} \left[
    1 + (-1)^\ell
    \right]
\end{eqnarray}
derived in Sec. \ref{s-251} [see Eq.~(\ref{a2l0})], reads
($\ell+m>0$):
\begin{eqnarray}
    a_{\ell,m}(0) =  \frac{(\ell+m-1)!}{2^{2(\ell+m)+3}}
    \left[
    \left(
        \frac{f_\beta}{2}+1
    \right)^m + (-1)^\ell \left(
        \frac{f_\beta}{2}-1
    \right)^m
    \right].
\end{eqnarray}
Combined with Eq.~(\ref{ans-joint}), it yields the leading term in
the $1/n$ expansion for joint cumulants of Landauer conductance and
the noise power:
\begin{eqnarray}
\label{ans-joint-answer}
    \delta\kappa_{\ell,m} \simeq \frac{1}{8}
    \frac{(\ell+m-1)!}{(4n)^{\ell+m}}
    \left[
    \left(
        \frac{f_\beta}{2}+1
    \right)^m + (-1)^\ell \left(
        \frac{f_\beta}{2}-1
    \right)^m
    \right].
\end{eqnarray}
Equations (\ref{klm-gauss}) and (\ref{ans-joint-answer}) are the
central result of this Section.

In particular, it brings an the following large-$n$ expression for
the cumulants of noise power in case of symmetric leads:
\begin{eqnarray} \label{npc} \fl
    \quad \langle\!\langle{\mathcal P}^\ell \rangle\!\rangle \simeq
    (G_0 \theta)^\ell \Bigg[ 2n \left(
    1+\frac{f_\beta}{4}
    \right) \delta_{\ell,1} + \left(
    1 + \frac{f_\beta^2}{8}
    \right) \delta_{\ell,2} \nonumber \\
    + \frac{(\ell-1)!}{8n^\ell}
    \left[
    \left( \frac{f_\beta}{2}-1
    \right)^\ell
    +
    \left( \frac{f_\beta}{2}+1
    \right)^\ell
    \right] \Bigg].
\end{eqnarray}
Explicit dependence of the noise power cumulants on both the
temperature ${\theta}$ and the bias voltage $\upsilon$ enters
through a single function $f_\beta$ [see Eq.~(\ref{fbeta})] that
depends on the ratio $\beta=\upsilon/2\theta$. Based on
Eq.~(\ref{npc}), it can further be shown that small but nonvanishing
cumulants of the third and higher order are responsible for long
exponential tails in the otherwise Gaussian distribution of the
noise power (compare with Vivo, Majumdar and Bohigas 2008).

\section{Conclusions}

In summary, we have presented an advanced formulation of the
recently proposed integrable theory of quantum transport (Osipov and
Kanzieper 2008) to study statistics of noise power fluctuations in a
chaotic cavity with broken time-reversal symmetry in the crossover
regime $(\theta,0)\rightarrow (0,\upsilon)$ between thermal and shot
noise. By {\it relating the cumulants of noise power to those of the
Landauer conductance}, we determined dependence of the noise power
cumulants (as well as of joint cumulants of Landauer conductance and
the noise power) on the bias voltage $\upsilon$, temperature
$\theta$, and the number of channels $N_{\rm L,R}$ in the leads
attached to a cavity through ballistic point contacts.

We are confident that ideas of integrability combined with the
scattering matrix approach are able to provide a nonperturbative
description of many more transport phenomena in chaotic cavities.
Quantum transport in cavities with losses (Doron, Smilansky and
Frenkel 1991, Beenakker and Brouwer 2001, Simon and Moustakas 2006)
and non-ideal leads (Brouwer 1995) are just two examples of chaotic
scattering systems whose detailed study is much called for.
\newline\newline
{\it Note added.}---Recently, we learnt about the paper by
Khoruzhenko {\it et al} (2009) where an alternative approach was
developed to describe statistics of conductance and shot-noise power
in chaotic cavities with and without time-reversal symmetry.
Possibly triggered by the earlier paper by Novaes (2008), these
authors combine a theory of the Selberg integral with the theory of
symmetric functions to evaluate the (joint) moments of Landauer
conductance and the shot-noise power in terms of series over all
partitions of the moment's order. In particular, Khoruzhenko {\it et
al} (2009) confirm our large-$n$ formulae Eqs.~(\ref{cum-res}) and
(\ref{npc}) at zero temperature.

\section*{Acknowledgements}
This work was supported by Deutsche Forschungsgemeinschaft SFB/Tr
12, and by the Israel Science Foundation through the grants No
286/04 and No 414/08.
\smallskip\smallskip\smallskip

\section*{Appendix}
\setcounter{section}{0}
\renewcommand{\thesection}{\Alph{section}}
\renewcommand{\theequation}{\Alph{section}.\arabic{equation}}

\section[Cumulant generating function for the Landauer conductance]{Cumulant generating
function for the Landauer conductance and the fifth Painlev\'e
transcendent} {\it Relation to a gap formation probability.}---The
``simplest'' (albeit not operative) way to observe that the
conductance cumulant generating function ${\mathcal F}_n(z,0)$
[Eq.~(\ref{ccgf})] can be expressed in terms of Painlev\'e V is to
spot that ${\mathcal F}_n(z,0)$ is essentially the gap formation
probability within the interval $(z,+\infty)$ in the spectrum of an
$n \times n$ Laguerre unitary ensemble. Indeed, the transformation
of integration variables $\lambda_j = z T_j$ in Eq.~(\ref{ccgf})
yields
\begin{eqnarray}
    {\mathcal F}_n(z,0) = c_n^{-1} z^{-n(n+\nu)}
    \int_{(0,z)^n} \prod_{j=1}^n d\lambda_j \lambda_j^\nu e^{-\lambda_j}\, \Delta_n^2({\bm \lambda}).
\end{eqnarray}
A nonperturbative evaluation of the above $n$-fold integral is
readily available (Tracy and Widom 1994, Forrester and Witte 2002)
eventually resulting in the following Painlev\'e V representation
(Osipov and Kanzieper 2008):
\begin{eqnarray}
\label{FnP}
    {\mathcal F}_n(z,0) = \exp\left(
        \int_0^z dt \frac{\sigma_n(t) - n(n+\nu)}{t}
    \right).
\end{eqnarray}
Here, $\sigma_n(t)$ satisfies the Jimbo-Miwa-Okamoto form of the
Painlev\'e V equation (Jimbo {\it et al} 1980, Okamoto 1987):
\begin{equation} \fl
\label{pv}
    (t \sigma_n^{\prime\prime})^2 + [\sigma_n - t \sigma_n^\prime
    + 2 (\sigma_n^\prime)^2 + (2n+\nu)\sigma_n^\prime]^2
    + 4(\sigma_n^\prime)^2 (\sigma_n^\prime + n) (\sigma_n^\prime+n+\nu)=0
\end{equation}
subject to the boundary condition $\sigma_{\rm V}(t\rightarrow
0)\simeq n(n+\nu)$. Keeping in mind the parameterisation
$n=\min(N_{\rm L},N_{\rm L})$ and $\nu = |N_{\rm L}-N_{\rm R}|$, one
concludes that Eq.~(\ref{pv}) is equivalent to Eq.~(\ref{s5-chazy})
announced in Sec. \ref{s-223}. \noindent\newline\newline {\it Direct
evaluation of ${\mathcal F}_n(z,0)$.}---To directly evaluate
${\mathcal F}_n(z,0)$ defined by Eq.~(\ref{ccgf}), we introduce the
associated $\tau$ function
\begin{eqnarray}
    \label{fnz0-tau-def}
    \tau_n({\bm t};z) = \frac{1}{n!}\int_{(0,1)^n} \prod_{j=1}^n dT_j\,T_j^\nu \,e^{-z T_j + V({\bm t};T_j)} \Delta_n^2({\bm T})
\end{eqnarray}
such that
\begin{eqnarray}
    {\mathcal F}_n(z,0) = \frac{n!}{c_n} \tau_n({\bm t}; z)\Big|_{{\bm t}={\bm 0}},
\end{eqnarray}
and make use of the KP equation Eq.~(\ref{fkp}),
\begin{eqnarray} \fl
    \label{fnz0-fkp}
        \left(
    \frac{\partial^4}{\partial t_1^4} + 3\,\frac{\partial^2}{\partial t_2^2} -
        4\, \frac{\partial^2}{\partial t_1 \partial t_3}
    \right)\, \log \tau_n({\bm t};z)
    + \,6\, \left(
        \frac{\partial^2}{\partial t_1^2}\, \log \tau_n({\bm t};z)
    \right)^2 = 0,
\end{eqnarray}
supplemented by the Virasoro
constraints~\footnote{Equations~(\ref{fnz0-vc-1}) and
(\ref{fnz0-vc-2}) readily follow from Eqs.~(\ref{vc-1}) and
(\ref{vc-2}) upon setting $f_\beta=0$ and $w=0$.}
\begin{equation}
\label{fnz0-vc-1}
    \big[ \hat{L}_{q+1}({\bm t}) - \hat{L}_{q}({\bm t})\big] \tau_n({\bm t};z)
     = 0,\;\;\;  q\ge 0,
\end{equation}
where a set of differential operators
\begin{eqnarray}
\label{fnz0-vc-2}
    \hat{L}_q({\bm t})
     =
     \hat{\mathcal L}_{q}({\bm t}) - z \frac{\partial}{\partial t_{q+1}}
        + \nu \frac{\partial}{\partial t_q}
\end{eqnarray}
involves the Virasoro operators Eq.~(\ref{vc-3}) [the convention
$\partial/\partial t_0 \equiv n$ is assumed].

In order to project the KP equation Eq.~(\ref{fnz0-fkp}) onto ${\bm
t}={\bm 0}$, we need only two Virasoro constraints labeled by $q=0$,
\begin{eqnarray} \fl \label{q0}
    \left[
    \sum_{j=1}^\infty jt_j \left(
    \frac{\partial}{\partial t_{j+1}} - \frac{\partial}{\partial t_{j}}\right)
    - z \frac{\partial}{\partial t_2} +(2n+\nu+z) \frac{\partial}{\partial t_1}
    \right] \log \tau_n({\bm t}; z)=n(n+\nu),
\end{eqnarray}
and $q=1$
\begin{eqnarray} \fl \label{q1}
    \Bigg[
    \sum_{j=1}^\infty jt_j \left(
    \frac{\partial}{\partial t_{j+2}} - \frac{\partial}{\partial t_{j+1}}\right)
    -z \frac{\partial}{\partial t_3} + (2n+\nu+z) \frac{\partial}{\partial t_2}\nonumber\\
     - (2n+\nu) \frac{\partial}{\partial t_1}
    + \frac{\partial^2}{\partial t_1^2}\Bigg] \log \tau_n({\bm t}; z) + \left(
    \frac{\partial}{\partial t_1} \log \tau_n({\bm t},z)
    \right)^2=0.
\end{eqnarray}
Here
\begin{eqnarray}
\frac{\partial}{\partial t_1} \tau_n({\bm t};z) = -
\frac{\partial}{\partial z} \tau_n({\bm t};z).
\end{eqnarray}
Lengthy but straightforward manipulations with Eqs.~(\ref{q0}) and
(\ref{q1}) as well as with their derivatives over $t_1$ and $t_2$
projected onto ${\bm t}={\bm 0}$ result in Eqs.~(\ref{sF}) and
(\ref{OK-2008}).

\section*{References}
\fancyhead{} \fancyhead[RE,LO]{References}
\fancyhead[LE,RO]{\thepage}

\begin{harvard}

\item[] Adler M and van Moerbeke P 1995
        Matrix integrals, Toda symmetries, Virasoro constraints, and orthogonal polynomials
        {\it Duke Math J} {\bf 80} 863

\item[] Adler M, Shiota T and van Moerbeke P 1995
        Random matrices, vertex operators and the Virasoro algebra
        {\it Phys. Lett. A} {\bf 208} 67

\item[] Adler M and van Moerbeke P 2001
        Hermitian, symmetric and symplectic random ensembles: PDE's for the distribution of the spectrum
        {\it Ann. Math.} {\bf 153} 149

\item[] Agam O, Aleiner I and Larkin A 2000
        Shot noise in chaotic systems: "Classical" to quantum crossover
        {\it Phys. Rev. Lett.} {\bf 85} 3153

\item[] Baranger H U and Mello P A 1994
        Mesoscopic transport through chaotic cavities: A random S-matrix theory approach
        {\it Phys. Rev. Lett.} {\bf 73} 142

\item[] Beenakker C W J 1997
        Random matrix theory of quantum transport
        {\it Rev. Mod. Phys.} {\bf 69} 731

\item[] Beenakker C W J and Brouwer P W 2001
        Distribution of the reflection eigenvalues of a weakly
        absorbing chaotic cavity
        {\it Physica E} {\bf 9} 463

\item[] Blanter Ya M and B\"uttiker M 2000
        Shot noise in mesoscopic Conductors
        {\it Phys. Rep.} {\bf 336} 1

\item[] Blanter Ya M and Sukhorukov E V 2000
        Semiclassical theory of conductance and noise in open chaotic cavities
        {\it Phys. Rev. Lett.} {\bf 84} 1280

\item[] Bl\"umel R and Smilansky U 1990
        Random-matrix description of chaotic scattering: Semiclassical approach
        {\it Phys. Rev. Lett.} {\bf 64} 241

\item[] Brouwer P W 1995
        Generalized circular ensemble of scattering matrices for a chaotic cavity with non-ideal leads
        {\it Phys. Rev. B} {\bf 51} 16878

\item[] B\"uttiker M 1990
        Scattering theory of thermal and excess noise in open conductors
        {\it Phys.~Rev.~Lett.} {\bf 65} 2901

\item[] B\"uttiker M 1992
        Scattering theory of current and intensity noise correlations in conductors and wave guides
        {\it Phys.~Rev.~B} {\bf 46} 12485

\item[] Chazy J 1911
        Sur les \'equations diff\'erentielles du troisi\`{e}me ordre et
        d'ordre sup\'erieur dont l'int\'egrale g\'en\'erale a ses
        points critiques fixes
        {\it Acta Math.} {\bf 34} 317

\item[] Cosgrove C M and Scoufis G 1993
        Painlev\'e classification of a class of differential equations
        of the second order and second degree,
        {\it Stud. Appl. Math.} {\bf 88} 25

\item[] Cron R, Goffman M F, Esteve D and Urbina C 2001
        Multiple-charge-quanta shot noise in superconducting atomic contacts
        {\it Phys. Rev. Lett.} {\bf 86} 4104

\item[] Date E, Kashiwara M, Jimbo M and Miwa T 1983
        Transformation groups for soliton equations, in:
        {\it Nonlinear Integrable Systems -- Classical Theory and
        Quantum Theory} edited by Jimbo M and Miwa T
        (World Scientific: Singapore)

\item[] Doron E, Smilansky U and Frenkel A 1991
        Chaotic scattering and transmission fluctuations
        {\it Physica D} {\bf 50} 367

\item[] Flindt C, Fricke C, Hohls F, Novotn\'y T, Neto\u{c}n\'{y} K, Brandes T and Haug R J 2009
        Universal oscillations in counting statistics
        Proc. Natl. Acad. Sci. USA {\bf 106} 10116

\item[] Forrester P J and Witte N S 2002
        Application of the $\tau$ function theory of Painlev\'e
        equations to random matrices: PV, PIII, the LUE, JUE and CUE
        {\it Commun. Pure Appl. Math.} {\bf 55} 679

\item[] Forrester P J 2006
        Quantum conductance problems and the Jacobi ensemble
        {\it J. Phys. A: Math. Gen.} {\bf 39} 6861

\item[] Hua L K 1963
        {\it Harmonic Analysis of Functions of Several Complex Variables in the Classical Domains}
        (American Mathematical Society, Providence)

\item[] Imry Y 2002
        {\it Introduction to Mesoscopic Physics} (Oxford University Press, New York)

\item[] Jimbo M, Miwa T, M\^ori Y and Sato M 1980
        Density matrix of an impenetrable Bose gas and the fifth
        Painlev\'e transcendent
        {\it Physica D} {\bf 1} 80

\item[] Kanzieper E 2002
        Replica field theories, Painlev\'e transcendents, and exact correlation functions
        {\it Phys. Rev. Lett.} {\bf 89} 250201

\item[] Kanzieper E 2005
        Exact replica treatment of non-Hermitean complex random matrices, in: {\it Frontiers in Field Theory}
        edited by Kovras O (Nova Science Publishers, New York)

\item[] Kanzieper E 2009
        Replica approach in random matrix theory {\it arXiv:~0909.3198} -- to appear in: {\it The Oxford Handbook of Random Matrix Theory} edited by Akemann G, Baik J and Di Francesco P (Oxford University Press, Oxford)

\item[] Khlus V A 1987
        Current and voltage fluctuations in micro-junctions of normal and superconducting metals
        {\it Sov.~Phys.~JETP} {\bf 66} 1243

\item[] Khoruzhenko B A, Savin D V and Sommers H-J 2009
        Systematic approach to statistics of conductance and shot-noise in chaotic cavities
        {\it Phys. Rev. B} {\bf 80}, 125301
        
\item[] Lesovik G B 1989
        Excess quantum noise in 2D ballistic point contacts
        {\it JETP Lett.} {\bf 49} 592

\item[] Lewenkopf C H and Weidenm\"uller H A 1991
        Stochastic versus semiclassical approach to quantum chaotic scattering
        {\it Ann. Phys. (N.Y.)} {\bf 212} 53

\item[] Martin T and Landauer R 1992
        Wave-packet approach to noise in multichannel mesoscopic systems
        {\it Phys.~Rev.~B} {\bf 45} 1742

\item[] Martin T 2005
         Noise in mesoscopic physics,
         in: {\it Nanophysics: Coherence and Transport} edited by Bouchiat H, Gefen Y, Gu\'eron S,
         Montambaux G, and Dalibard J (Elsevier, Amsterdam)

\item[] Mehta M L 2004
        {\it Random Matrices} (Amsterdam: Elsevier)

\item[] Mello P A and Baranger H U 1999
        Interference phenomena in electronic transport through chaotic cavities: An information-theoretic approach
        {\it Waves Random Media} {\bf 9} 105

\item[] Mironov A and Morozov A 1990
        On the origin of Virasoro constraints in matrix models: Lagrangian approach
        {\it Phys. Lett. B} {\bf 252} 47

\item[] Morozov A Yu 1994
        Integrability and matrix models
        {\it Physics-Uspekhi} {\bf 37} 1

\item[] M\"uller S, Heusler S, Braun P and Haake F 2007
        Semiclassical approach to chaotic quantum transport
        {\it New J. Phys.} {\bf 9} 12

\item[] Novaes M 2008
        Statistics of quantum transport in chaotic cavities with
        broken time-reversal symmetry
        {\it Phys. Rev. B} {\bf 78} 035337

\item[] Oberholzer S, Sukhorukov E V, Strunk C, Sch\"onenberger C, Heinzel T and Holland M 2001
        Shot noise by quantum scattering in chaotic cavities
        {\it Phys. Rev. Lett.} {\bf 86} 2114

\item[] Oberholzer S, Sukhorukov E V and Sch\"onenberger C 2002
        Crossover between classical and quantum shot noise in chaotic cavities
        {\it Nature} {\bf 415} 765

\item[] Okamoto K 1987
        Studies on the Painlev\'e equations, II. Fifth Painlev\'e equation PV
        {\it Jpn. J. Math.} {\bf 13} 47

\item[] Osipov V Al and Kanzieper E 2007
        Are bosonic replicas faulty?
        {\it Phys. Rev. Lett.} {\bf 99} 050602

\item[] Osipov V Al and Kanzieper E 2008
        Integrable theory of quantum transport in chaotic cavities
        {\it Phys. Rev. Lett.} {\bf 101} 176804

\item[] Osipov V Al and Kanzieper E 2009
        Correlations of RMT characteristic polynomials and integrability: I.~Hermitean matrices
        {\it manuscript}

\item[] Politzer H D 1989
        Random-matrix description of the distribution of mesoscopic
        conductance
        {\it Phys. Rev. B} {\bf 40} 11917

\item[] Richter K and Sieber M 2002
        Semiclassical theory of chaotic quantum transport
        {\it Phys. Rev. Lett.} {\bf 89} 206801

\item[] Savin D V and Sommers H J 2006
        Shot noise in chaotic cavities with an arbitrary number of open channels
        {\it Phys. Rev. B} {\bf 73} R081307

\item[] Savin D V, Sommers H-J and Wieczorek W 2008
        Nonlinear statistics of quantum transport in chaotic cavities
        {\it Phys. Rev. B} {\bf 77} 125332

\item[] Simon S H and Moustakas A L 2006
        Crossover from conserving to lossy transport in circular
        random matrix ensembles
        {\it Phys. Rev. Lett.} {\bf 96} 136805

\item[] Tracy C A and Widom H 1994
        Fredholm determinants, differential equations and matrix
        models
        {\it Commun. Math. Phys.} {\bf 163} 33

\item[] Tu M H, Shaw J C and Yen H C 1996
        A note on integrability in matrix models
        {\it Chinese J. Phys.} {\bf 34} 1211

\item[] Vivo P, Majumdar S N and Bohigas O 2008
        Distributions of conductance and shot noise and associated phase transitions
        {\it Phys. Rev. Lett.} {\bf 101} 216809

\smallskip
\end{harvard}

\end{document}